\documentclass[a4paper,10pt,twoside]{cpc-hepnp}

\usepackage{multicol}
\usepackage{graphicx}
\usepackage{booktabs}
\usepackage{amssymb,bm,mathrsfs,amscd}
\usepackage[tbtags]{amsmath}
\usepackage{lastpage}

\usepackage{epsfig}
\usepackage{mathptmx}      
\usepackage{latexsym}
\usepackage{comment}
\usepackage{mathtools}

\usepackage{amsfonts}
\usepackage{hyperref}
\usepackage{color}

\usepackage{supertabular} 
\usepackage{epstopdf}

\usepackage[english]{babel}
\usepackage{graphics}
\usepackage{placeins}
\usepackage{epsfig}
\usepackage{diagbox}

\usepackage{soul}

\begin{document}

\fancyhead[c]{\small Chinese Physics C~~~Vol. xx, No. x (xxxx)
xxxxxx} \fancyfoot[C]{\small xxxxxx-\thepage}


\title{Is the $X(3872)$ a bound state ?}

\author{%
      Pablo G. Ortega$^{1)}$\email{pgortega@usal.es}%
\quad Enrique Ruiz Arriola$^{2)}$\email{earriola@ugr.es}
}
\maketitle

\address{%
$^1$ Grupo de F\'isica Nuclear and Instituto Universitario de F\'isica 
Fundamental y Matem\'aticas (IUFFyM), Universidad de Salamanca, E-37008 
Salamanca, Spain\\
$^2$ Departamento de
  F\'{\i}sica At\'omica, Molecular y Nuclear \\ and Instituto Carlos I
  de F{\'\i}sica Te\'orica y Computacional, Universidad de Granada,
  E-18071 Granada, Spain.\\
}

\begin{abstract}
All existing experimental evidence of the bound state nature of the
$X(3872)$ relies on considering its decay products with a finite
experimental spectral mass resolution which is typically $\Delta m \ge
2 $MeV and much larger than its alleged binding energy,
$B_X=0.00(18)$MeV.  On the other hand, we have found recently that
there is a neat cancelation in the $1^{++}$ channel for the invariant
$D \bar D^*$ mass around the threshold between the continuum and bound
state contribution. This is very much alike a similar cancelation in
the proton-neutron continuum with the deuteron in the $1^{++}$
channel. Based on comparative fits of experimental cross section
deuteron and $X(3872)$ prompt production in pp collisions data with a
finite $p_T$ to a common Tsallis distribution we find a strong
argument questioning the bound state nature of the state but also
explaining the large observed production rate likely consistent with a
half-bound state.
\end{abstract}

\begin{keyword}
Charmonium molecular states, Particle Production, Tsallis distribution
\end{keyword}

\begin{pacs}
12.39.Pn, 14.40.Lb, 14.40.Rt
\end{pacs}


\begin{multicols}{2}

\section{Introduction}

The early possibility of loosely bound states near the charm threshold
first envisaged in Ref.~\cite{Nussinov:1976fg} seems to be confirmed
now by the wealth of evidence on the existence of the $X(3782)$ state
with binding energy $B_X=M_D+M_{\bar D^*}
-M_X=0.00(18)$MeV~\cite{Tanabashi:2018oca} and which has triggered a
revolution by the proliferation of the so-called X,Y,Z states (for
reviews see e.g.~\cite{Esposito:2016noz,Guo:2017jvc}). In the absence
of electroweak interactions this state has the smallest known hadronic
binding energy.  However, since this state is unstable, all the
detection methods of the $X(3872)$ are based on looking for its decay
channels spectra such as $X\to J/\psi\pi^+\pi^-$ where the mass
resolution never exceeds $\Delta m \sim 1-2
$MeV~\cite{Choi:2003ue,Aubert:2004zr,Choi:2011fc,Aaij:2013zoa} (see
e.g. \cite{Karliner:2017qhf} for a pictorial display on spectral
experimental resolution). Therefore it is in principle unclear if one
can determine the mass of the $X(3872)$ or equivalently its binding
energy $\Delta B_X \ll \Delta m$ with such a precision, since we
cannot distinguish sharply the initial state.

In most analyses up to now (see however \cite{Kang:2016jxw}) the bound
state nature is assumed rather than deduced. In fact, the molecular
interpretation has attracted considerable attention, since for a
loosely bound state many properties are mainly determined by its
binding energy~\cite{Guo:2017jvc} and characterized by a 
line shape in production processes~\cite{Guo:2019qcn}.  However, we
have noticed recently a neat and accurate cancellation between the
would-be X(3872) bound state and the $D \bar D^*$ continuum which has
a sizable impact on the occupation number at finite
temperature~\cite{Ortega:2017hpw,Ortega:2017shf}.  This reduction
stems from a cancellation in the density of states in the $1^{++}$
channel and potentially blurs any detected signal where a
superposition of $1^{++}$ states is at work. Such a circumstance makes
us questioning in the present letter the actual character of the
state. We will do so by analyzing the $p_T$ distribution of the
$X(3872)$ in high energy production experiments and folding the
expected distribution with the actual mass distribution corresponding
to the $1^{++}$ spectrum via the level density within the accessible
experimental resolution. For our argument a qualitative and
quantitative comparison with a truely weakly bound state such as the
deuteron, $d$, will be most enlightening. As a matter of fact, the
similarities between $d$ and $X(3872)$ have been
inspiring~\cite{Tornqvist:1993ng,Close:2003sg,Braaten:2003he}. Compared
to the $X(3872)$ the main difference is that the deuteron is detected
{\it directly} by analyzing its well defined track and/or stopping
power. Actually, the production of loosely bound nuclei and
anti-nuclei, including $d$,$\bar d$, $^3{\rm He}_\Lambda$, etc. in
ultra-high pp collisions is a remarkable and surprissing experimental
observation in recent years~\cite{Braun-Munzinger:2018hat} and so far
poorly understood~\cite{Cai:2019jtk}.

The cancellation echoes a similar effect on the deuteron pointed out
by Dashen and Kane in their discussion on the counting of states in
the hadron spectrum in a coarse grained sense~\cite{Dashen:1974ns}
which we review in some detail in the next section. In section 3 we
analyze the consequences in a production process. Finally, in section
4 we draw our the conclusions and provide an outlook for future work.

\section{Dashen and Kane cancellation mechanism}

In order to illustrate the Dashen-Kane mechanism~\cite{Dashen:1974ns}
we introduce the cumulative number of states with invariant CM mass
$\sqrt{s}$ below $M$ in a given channel with fixed $J^{PC}$ quantum
numbers. This involves the $J^{PC}$ spectrum which contains bound
states and continuum states with threshold $M_{\rm th}$ and is given
as
\begin{eqnarray}
N(M)= \sum_i \theta(M-M_i^B)
  + \frac1\pi \sum_{\alpha=1}^{n} [\delta_\alpha (M)-\delta_\alpha (M_{\rm th})] \, .
\label{eq:ncum} 
\end{eqnarray}
where the index $i$ runs over the $M_i^B$ bound states and $\alpha$
over the $n$ coupled channels. Here we have separated bound states
$M_n^B$ explicitly from scattering states written in terms of the
eigenvalues of the S-matrix, i.e.  $S = U {\rm Diag} (e^{2i\delta_1},
\dots, e^{2i\delta_n} ) U^\dagger$, with $U$ a unitary transformation
for $n$-coupled channels and $\delta_i(M)$ the eigenphaseshifts for
the channel $i$ at CM invariant mass $\sqrt{s}=M$. This definition
fulfills $N(0)=0$. In the single channel case, and in the limit of
high masses $M \to\infty$ one gets $ N(\infty)=n_B + \frac1{\pi}
[\delta(\infty)-\delta(M_{\rm th})]=0 $ due to Levinson's theorem.
While the origin of the bound state term is quite obvious, the
derivation of the continuum term is a bit subtle but standard and can
be found in many textbooks on statistical mechanics dealing with the
quantum virial expansion (see
e.g.~\cite{Parisi:1988nd,huang2009introduction}). In potential
scattering it can be best deduced by confining the system in a large
spherical box which quantizes the energy and relates the energy shift
due to the interaction to the phase-shift and then letting the volume
of the system go to infinity~\cite{Dashen:1974ns}.

In the particular case of the deuteron, which is a neutron-proton
$1^{++}$ state bound by $B_d=2.2 {\rm MeV}$, the cancellation between
the continuum and discrete parts of the spectrum was pointed out by
Dashen and Kane long ago~\cite{Dashen:1974ns}.   (see
also \cite{Arriola:2015gra,Arriola:2014bfa} for an explicit picture
and further discussion within the resonance gas model framework). The
opening of new channels and the impact of confining interactions was
discussed in Ref.~\cite{Dashen:1976cf}. In the $1^{++}$ channel, the
presence of tensor force implies a coupling between the $^3S_1$ and
$^3D_1$ channels. While the partial wave analysis of NN scattering
data and the determination of the corresponding phase-shifts is a well
known subject~\cite{Perez:2013jpa}, we note that a similar analysis in
the $D \bar D^*$ case is at present in its infancy. In our first model
determination in Ref.~\cite{Ortega:2017hpw,Ortega:2017shf} the mixing
has an influence for larger energies than those considered here.
Therefore, in order to illustrate how the cancellation comes about, we
consider a simple model which works sufficiently accurately for {\it
  both} the deuteron and the $X(3872)$ by just considering a contact
(gaussian) interaction~\cite{Gamermann:2009uq} in the $^3S_1$-channel
and using efective range parameters to determine the corresponding
phase-shift in the $d$ and $X(3872)$
channels~\cite{Arriola:2013era,Ortega:2017hpw} respectively.

The result for $N(M)$ in both $d$ or $X$ cases depicted in
Fig.~\ref{fig:1} display a similar pattern for the $np$ or $D\bar D^*$
invariant masses respectively. The sharp rise of the cumulative number
is followed by a strong decrease generated by the phase-shift. For
larger invariant masses $M$ several effects appear, and in particular
the nuclear core (see e.g.~\cite{Arriola:2014bfa}) or composite nature
of the $X(3872)$ and its $c \bar c$ content becomes manifest (see
eg.~\cite{Ortega:2009hj}).

\begin{center}
\includegraphics[width=0.45\textwidth]{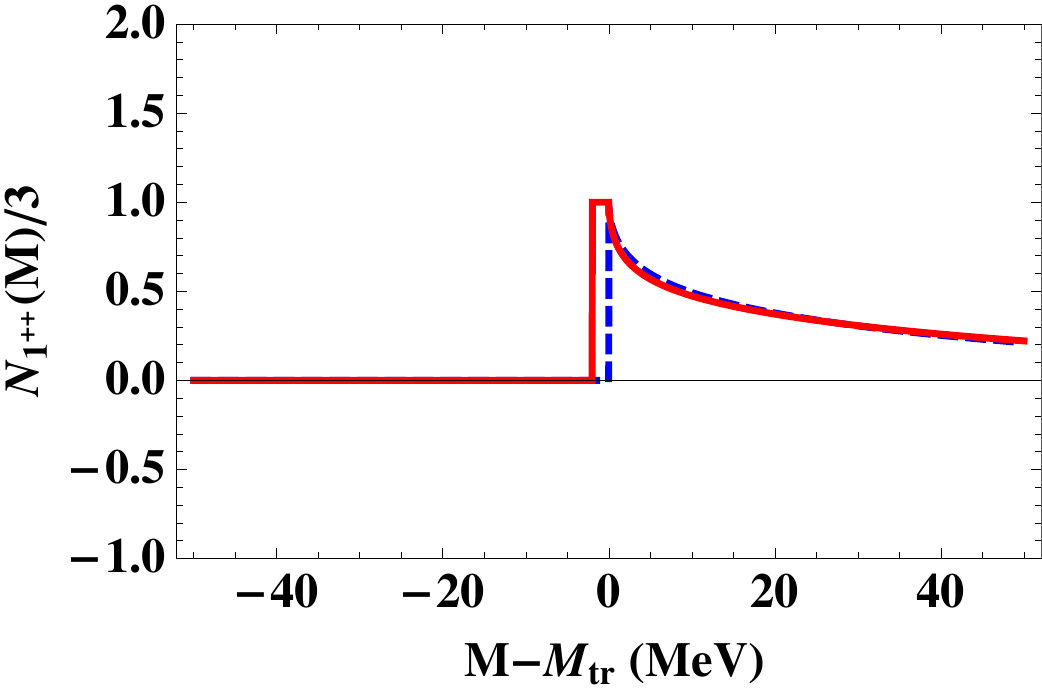}
\figcaption{\small{Cumulative number in the $1^{++}$ channel for the
    deuteron (solid) and $X(3872)$ (dashed) as a function of the
    invariant mass $M$ respect to np and $D \bar D^{*}$ values
    respectively. We divide by spin degeneracy.}\label{fig:1}}
\end{center}

An immediate consequence of this effect trivially follows from
Eq.~\eqref{eq:ncum} for an observable depending on the invariant mass
function $O(M)$.  The corresponding measured quantity for a bin in the
range $(m-\Delta m/2 , m+\Delta m/2)$ becomes
\begin{eqnarray}
  O_{\Delta m} \equiv \int_{m-\Delta m/2}^{m+\Delta m/2}  dM \rho(M) O (M) \, .
\label{eq:Obs}
\end{eqnarray}
where $\rho(M)$ is the density of states, defined as
\begin{eqnarray}
  \rho(M) = \frac{d N(M)}{dM}=
\sum_i \delta(M-M_i^B)
  + \frac1\pi \sum_{\alpha=1}^{n} \delta'_\alpha (M) \, ,
\label{eq:dndm}
\end{eqnarray}
where $\delta'_\alpha(M)$ denotes the derivative of the phase shift
with respect to the mass.

In the single channel case, with phase shift $\delta_\alpha (M)$, and
if the resolution is much larger than the binding energy $\Delta m \gg
|B| \equiv |M_B-M_{\rm tr}|$ one has
\begin{eqnarray}
  O|_{M^B \pm \Delta m} = O(M^B) 
  + \frac1\pi \int_{M_{\rm tr}}^{M_{\rm tr}+\Delta m/2}  dM \delta_\alpha ' (M) O(M)\, .
\label{eq:Obs-can}
\end{eqnarray}
which, on view of Fig.~\ref{fig:1} and for a smooth observable $O(M)$,
points to the cancellation, anticipated by Dashen and
Kane~\cite{Dashen:1976cf}. The effect was explicitly seen in the $np$
virial coefficient at astrophysical temperatures, $T \sim 1-10$ 
MeV~\cite{Horowitz:2005nd}. We have recently
shown~\cite{Ortega:2017hpw} how this cancellation can likewise be
triggered for the $X(3872)$ ocupation number at quark-gluon crossover
temperatures $T \sim 100-200$MeV. This will be relevant in
relativistic heavy ion collisions when X-production yields are
measured, because the partition function involves a Boltzmann factor,
$\sim e^{-\sqrt{p^2+m^2}/T}$ with the density of states,
Eq.~\eqref{eq:dndm} and the measured yields reproduce remarkably the
predictions occupation numbers in the hadron resonance gas
model~\cite{Andronic:2017pug}.

Therefore, given these tantalizing similarities a comparative study of
the deuteron {\it and} $X(3872)$ production rates to ultra-high
energies in colliders provides a suitable callibration tool in order
to see the effects of the Dashen-Kane cancellation due to the finite
resolution $\Delta m$ of the detectors signaling the $X(3872)$ state
via its decay products and decide on its bound state character. Here
we propose to study the effect in the observed tranverse momentum
($p_T$) distributions.

\section{$X(3872)$ production abundance}

While the theory regarding the shape of transverse momentum
distribution is not fully developped (see
e.g. Ref.~\cite{Sivers:1975dg} for an early review,
Ref.~\cite{Rak:2013yta} for a historical presentation), we will rest
on phenomenological ansatze which describe the data.  On the one hand,
the asymptotic $p_T$-spectrum \cite{Feynman:1978dt} provides a
production rate $1/p_T^8$ based on quark-quark scattering. Hagedorn
realized that an interpolation between the power correction and a
thermal Boltzmann $p_T$-distribution would
work~\cite{Hagedorn:1983wk}. A thermodynamic interpretation for
non-extensive systems~\cite{feshbach1987small} of the rapidity
distribution was proposed by Tsallis~\cite{Tsallis:1987eu} and first
applied to high energy phenomena in
Refs.~\cite{Bediaga:1999hv,Alberico:1999nh}, namely the differential
occupation number is given by
\begin{equation}
 \frac{d^3 N}{d^3 p}= \frac{g
   V}{(2\pi)^3}\left(1+(q-1)\frac{E(p)}{T}\right)^{-\frac{q}{q-1}} \xrightarrow{q\to 1}
 \frac{g V}{(2\pi)^3} e^{-\frac{E(p)}{T}}
\label{eq:dNd3p}
\end{equation}
where $E(p)=\sqrt{p^2+m^2}$, $V$ is volume of the system,
$T$ the temperature and $g$ the
degrees of freedom and, as indicated, the limit $q \to 1 $ produces
the Boltzmann distribution. We use here the form obtained by the
maximum Tsallis entropy principle~\cite{Megias:2015fra}.

The invariant differential production rate, $d^3N/(d^2 p_Tdy) \equiv
E_p d^3 N /d^3 p $ with $ y = \tanh^{-1}(E_p/p_z)$ the rapidity, has
the asymptotic matching corresponds to
$q=1.25$~\cite{Bhattacharyya:2017cdk}. While the thermodynamic
interpretation is essential to link the degrees of freedom $g$ with
the production rate~\cite{Cleymans:2012ya}, we note that we have
checked~\cite{ortega-arriola-2019} that a Tsallis distribution
describes accurately the results from particle Monte Carlo generators
such as PYTHIA~\cite{Sjostrand:2007gs,Sjostrand:2006za}. This
distribution has also been applied recently by the ALICE collaboration
to $d$-production~\cite{Acharya:2017fvb} in $pp$ collisions.

We show next that the $X(3872)$, $\Psi(2S)$ and deuteron prompt
production cross sections can be described with the {\it same} Tsallis
distribution:
\begin{equation}\label{ec:tsallis}
\frac1{2\pi p_T} \frac{d\sigma (m)}{dp_T}={\cal N}\int dy\,\, E(p_T,y) 
\left[1+ \frac{q-1}{T}E(p_T,y)\right]^{\frac{q}{1-q}}
\end{equation}
with $E(p_T,y)=\sqrt{p_T^2+m^2} \cosh y $ and ${\cal N}$ a
normalization factor. Obviously, a direct
comparison requires similar $p_T$ values as possible for both $d$,
$\Psi(2S)$ and $X(3872)$; the closest ones come from
ALICE~\cite{Acharya:2017fvb} and
CMS~\cite{Chatrchyan:2011kc,Chatrchyan:2013cld} respectively. The
ATLAS data for $X(3872)$~\cite{Aaboud:2016vzw} confirm a power law
behaviour in $p_T$ but extend over a much larger range than the
available $d$ and hence are not used in this study. The deuteron data
is given in invariant differential yields $d^2N/(2\pi p_Tdp_Tdy)$,
hence the inelastic $pp$ cross section at $\sqrt{s}=7$ TeV,
$\sigma_{\rm inel}^{pp}=73.2\pm1.3$ mb, as measured by
TOTEM~\cite{Antchev:2013iaa} has been used to transform it into
differential cross section.

\begin{center}
\tabcaption{ \label{tab:fit}\small{ Best fit of parameters for Tsallis distribution.  The X data used from CMS~\cite{Chatrchyan:2013cld} is multiplied by the branching fraction ${\cal B}_X\equiv{\cal B}(X\to J/\psi\pi^+\pi^-) $. 
Correlation between $q$ and $T$ parameters is practically $-1$ ($r=-0.9992$).}}
\footnotesize
\begin{tabular*}{80mm}{c|cc}
\toprule & $X(3872)+d$ &$X(3872)+\Psi(2S)+d$ \\
\hline
$\ln({\cal N}_X{\cal B}_X)$ & $41.4\pm0.4$ & $41.4\pm0.4$ \\
$\ln({\cal N}_d)$ & $40.35\pm0.09$ & $40.35\pm0.09$  \\
$\ln({\cal N}_{\Psi})$ & - & $44.3\pm0.2$ \\
$q$ & $1.122\pm0.001$ & $1.122\pm0.001$  \\
$T$ [MeV] & $7.017\pm0.07$ & $7.018\pm0.07$ \\
\hline $N_d$  & $(2.02\pm0.02)\cdot 10^{-4}$ & $(2.01\pm0.02)\cdot 10^{-4}$ \\
$N_X{\cal B}_X$  & $(9\pm 3)\cdot 10^{-6}$ & $(9\pm3)\cdot 10^{-6}$ \\
$N_{\Psi}$  & - & $(2.2\pm0.3)\cdot 10^{-4}$  \\
\hline
$\langle p_T \rangle_d$ & $1.102\pm0.007$ & $1.102\pm0.007$ \\
$\langle p_T \rangle_X$ & $2.249\pm0.015$ & $2.249\pm0.015$ \\
$\langle p_T \rangle_{\Psi}$ & $2.142\pm0.014$ & $2.142\pm0.014$ \\
\hline
$N_X{\cal B}_X/N_d$ & $0.046^{+0.016}_{-0.013}$ & $0.046^{+0.015}_{-0.013}$ \\
$N_{\Psi}/N_d$ & - & $1.09^{+0.16}_{-0.17}$ \\
\bottomrule
\end{tabular*}
\end{center}

On a phenomenological level we perform two fits: One including $d$ and
$X(3872)$ data and another one adding the $\Psi(2S)$ data. In both
cases the ${\cal N}_{d,X,[\Psi]}$, $q$ and $T$ are fitted by
minimizing the corresponding $\chi^2$ function with
\texttt{Minuit}~\cite{James:1975dr}. The experimental error in the
$x$-axis has been incorporated in the $\chi^2$ via a MonteCarlo
procedure with $5000$ runs, where the $p_T$ value of each experimental
data has been randomly shifted within the experimental range with an
uniform distribution. Due to the scarcity of X data, we assume the
production rate is mainly driven by the deuteron. That way, an initial
minimization of the $q$, $T$ and ${\cal N}_d$ is done, and the
resulting best-fit values of $q$ and $T$ are employed to fix ${\cal
  N}_X$ and ${\cal N}_{\Psi}$.

\begin{center}
\includegraphics[width=0.45\textwidth]{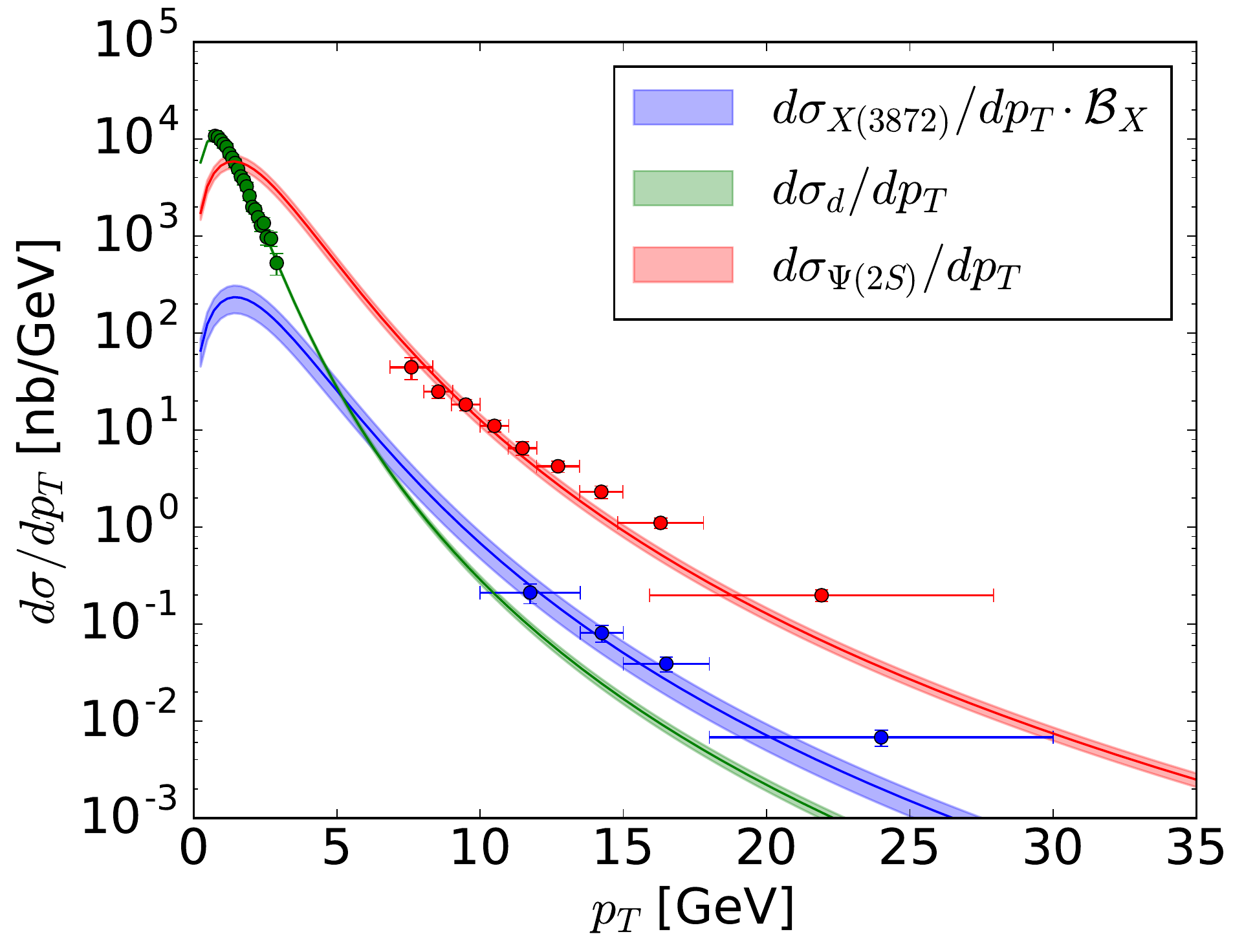}
\figcaption{\small{ Comparison between the prompt production cross
    section in $pp$ collisions of $X(3872)$ (blue), the deuteron
    (green) and the $\Psi(2S)$ (red). $\Psi(2S)$ data from CMS~\cite{Chatrchyan:2011kc}. 
    The $X(3872)$ data from CMS~\cite{Chatrchyan:2013cld} is multiplied by the branching
    fraction ${\cal B}(X\to J/\psi\pi\pi)$. Deuteron data in $pp$
    collisions are taken from ALICE~\cite{Acharya:2017fvb}. The lines are
    Tsallis distributions fitted to each data set, with the same $q$
    and $T$ parameters.  We fit both $X(3872)$, $\Psi(2S)$ and
    deuteron data. The shadowed bands represent the statistical $68\%$ confidence level (CL) obtained from the fit.}\label{fig:tsallis}}
\end{center}
\begin{center}
\includegraphics[width=0.45\textwidth]{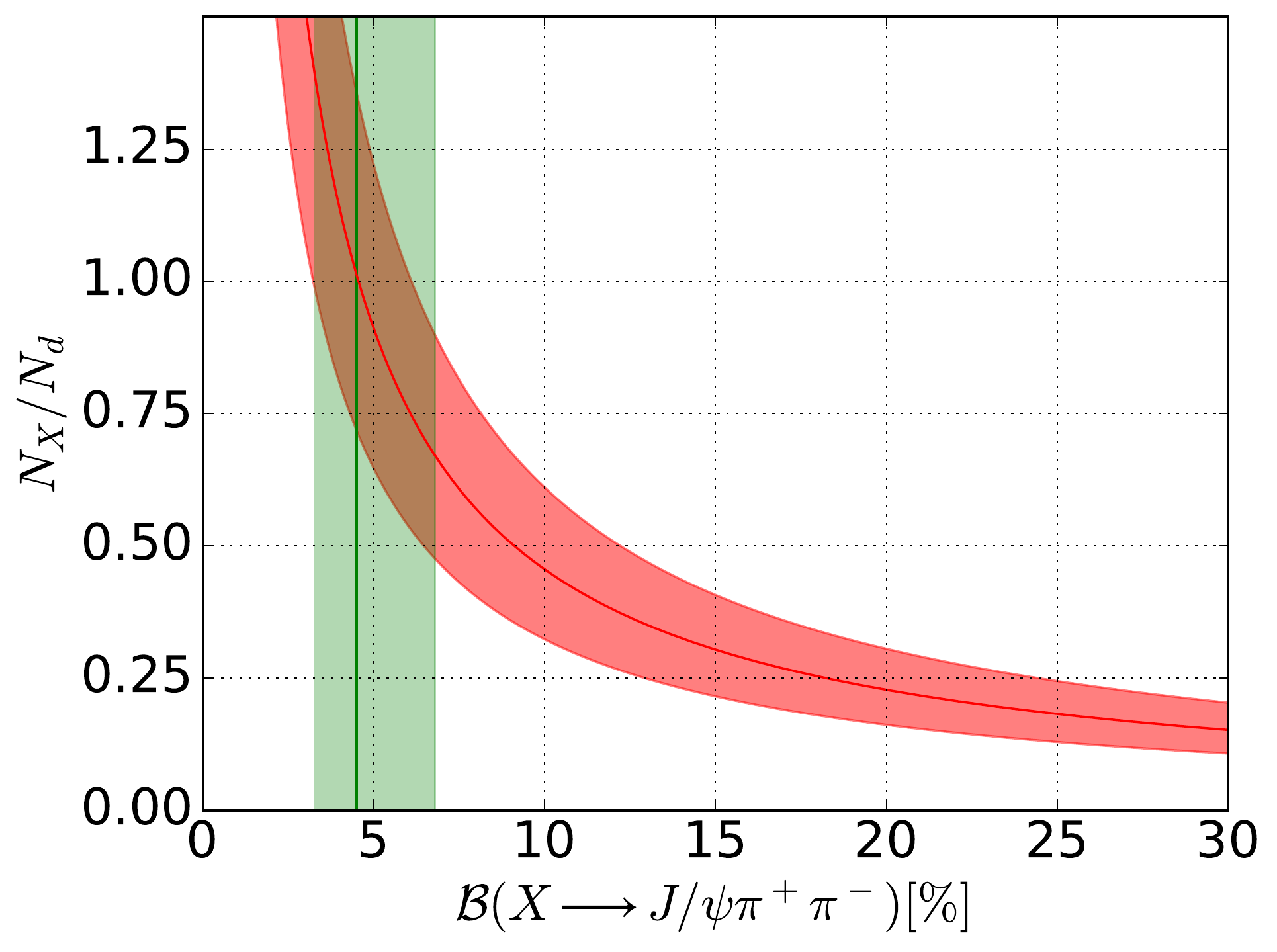}
\figcaption{\small{In red, the $X(3872)$ vs deuteron prompt production
    ratio as a function of the branching fraction ${\cal B}(X\to
    J/\psi\pi^+\pi^-)$ when fitting to $X(3872)$, $\Psi(2S)$ and
    deuteron data. The shadowed band represent the statistical $68\%$
    confidence level (CL) obtained from the fit. The green band shows
    the constrains of the recent analysis of C.~Li {\it et al} from
    BESIII data~\cite{ChangzhengYuan2019} $ {\cal B}_X\equiv{\cal
      B}(X\to J/\psi\pi^+\pi^-)=4.5^{+2.3}_{-1.2}$.
  }\label{fig:branching}}
\end{center}

The results can be found in Tab.~\ref{tab:fit}, and the final
production fit at Fig.~\ref{fig:tsallis} for the two considered
calculations, one with the $X(3872)$ and the deuteron, and another for
the $X(3872)$, the $\Psi(2S)$ and the deuteron. They are both
compatible, as expected, as the $X$ to $\Psi(2S)$ production ratio,
measured at CMS, is almost constant~\cite{Chatrchyan:2013cld}.  The
$X/d$ production ratio is $0.046^{+0.016}_{-0.013}$ for $X+d$ fit (and
practically the same value for X+$\Psi$+d fit), dependent on the value
of the branching fraction.  Note that we do not have the pure cross
section for the $X$, as it is multiplied by the unmeasured branching
fraction which has been recently constrained in an analysis of BESIII
data by C.~Li {\it et al}~\cite{ChangzhengYuan2019} to be $ {\cal
  B}_X\equiv{\cal B}(X\to J/\psi\pi^+\pi^-)=4.5^{+2.3}_{-1.2}
\%$. This value is consistent with the PDG lower- ${\cal
  B}_X>3.2\%$~\cite{Patrignani:2016xqp}, and upper bound ${\cal
  B}_X<6.6\%$ at $90\%$ C.L.~\cite{Yuan:2009iu}. The uncertainty comes
from the most recent value of ${\cal B}_X\cdot(B^-\to K^- X(3872))<
2.6\times 10^{-4}$ at $90\%$ C.L.~\cite{Kato:2017gfv}. We note that in
a recent paper, Esposito {\it et al.}~\cite{Esposito:2015fsa} consider
the wider range $8.1^{+1.9}_{-3.1}\%$.

Consequently, we can study the ratio of the X/d occupation numbers as
a function of the ${\cal B}_X$ branching fraction. In
Fig.~\ref{fig:branching} we see the results. Considering the error
bars, the experimental constrains give ratios between $0.3$ and $1.9$
for $N_X /N_d$.

In our previous fits above we have neglected the role played by the
finite resolution of the detectors, $\Delta m$, which we dicuss next.
Ref.~\cite{Chatrchyan:2013cld} uses a $\pm 2\sigma$ window around the
$X(3872)$ mass, with $\sigma=5-6$ MeV, to select the $X(3872)$ events
in the $J/\psi\pi\pi$ invariant mass spectrum. That means that the
branching fraction ${\cal B}(X\to J/\psi\pi^+\pi^-)$, as measured by
CMS is averaged in the $\left[M_X-2\sigma,M_X+2\sigma\right]$ energy
window, which includes the continuum.  As a consequence of the energy
window, there are many decays that can be affected, those involving
the $\bar D^0D^{0\,*}$ channel.

\begin{center}
  \includegraphics[width=0.45\textwidth]{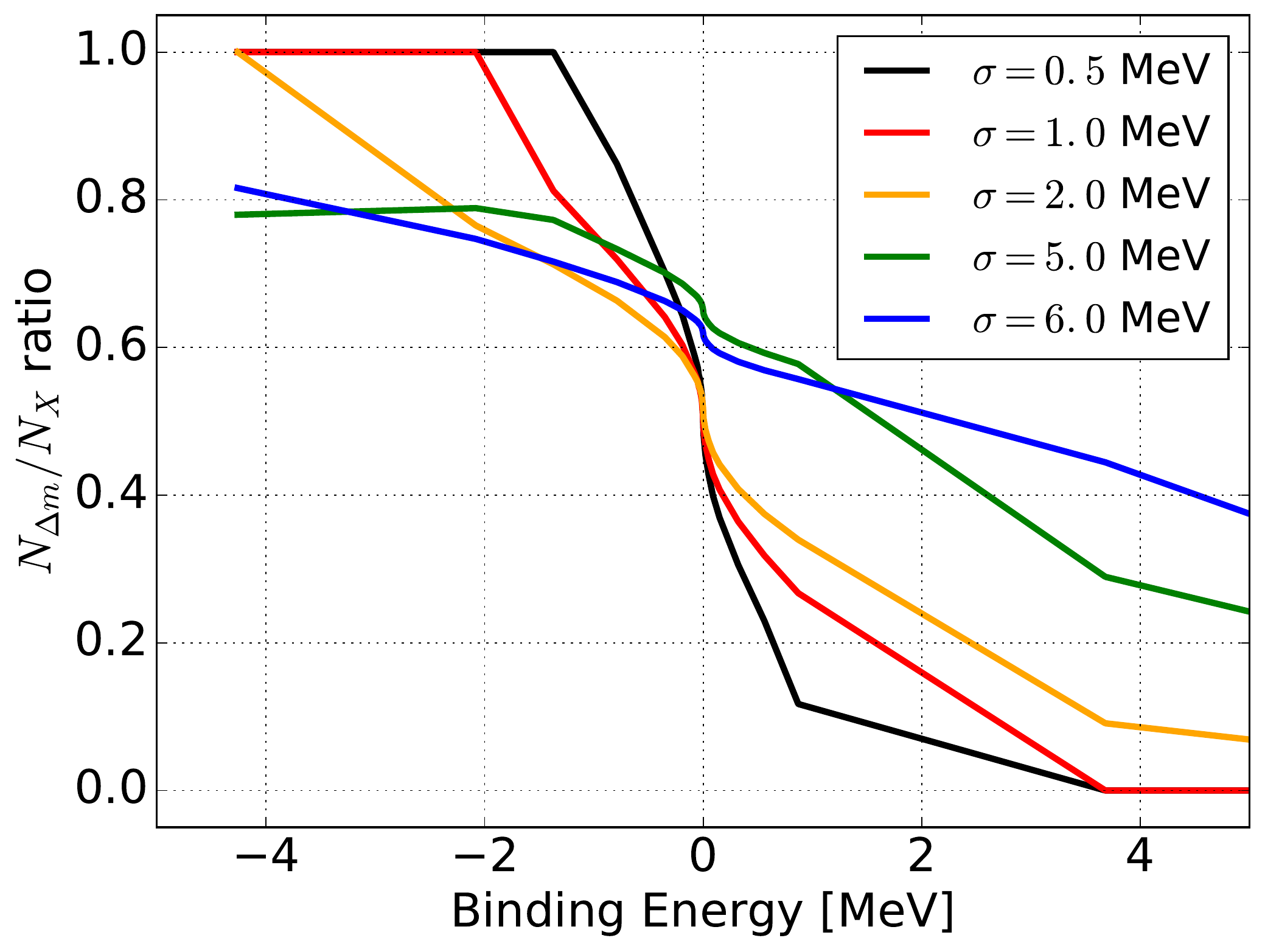}
\figcaption{\small{The relative occupation number {\it with} and {\it
        without} finite resolution $\Delta m= 2 \sigma$ as a function
      of the $X(3872)$ binding energy. Using $M_X = M_D + M_{\bar D^*}
      - \gamma_{X}^2/(2 \mu_{D , \bar D^*})$ we take the positive
      branch for the virtual state ($\gamma_X > 0$) and the negative
      branch for the bound state ($\gamma_X < 0$) for different
      $\sigma$ values.}\label{fig:tsallis-factor}}
\end{center}

In fact, the distribution obtained from Eq.~\ref{eq:dNd3p} {\it
  depends} on the mass, and hence its observed value undergoes
formula~\ref{eq:Obs-can}, reflecting the finite resolution.  Similar
to the finite temperature case~\cite{Ortega:2017hpw} we have checked
that the Tsallis $p_T$-shape is basically preserved for $p_T \gg
\Delta m$, but the occupation number is modified for $\Delta m \gg B
$.

For definiteness we use $\Delta m=2\sigma$, as CMS measures the
$X(3872)$ in a $\pm2\sigma$ region around the central value of the X
mass. The net effect is summarized in a ratio, which we find to be
practically independent of the transverse momentum $p_T$ for the
Tsallis distribution shape,
\begin{equation}
  \frac{\sigma_{m=M_X \pm \Delta m }}{\sigma_{M_X}} \sim \frac{{\cal N}_{\Delta m}}{{\cal N}_X} \, . 
\end{equation}
This formula will allow us to set values for the {\it relative}
occupation numbers due to the finite resolution. We take $M_X = M_D +
M_{\bar D^*} - \gamma_{X}^2/(2 \mu_{D , \bar D^*})$ as a parameter by
looking at the poles of the $D \bar D^*$ S-matrix in the $^3S_1-^3D_1$
channel~\cite{Ortega:2017hpw}.  Therefore, while in the limit of
$\Delta m \to 0$ we should expect the ratio $N_X/N_d \to 1$, $1/2$ or
$0$ for a bound ($\gamma_X > 0$), half-bound ($\gamma_X = 0$) or
unbound (actually virtual, ($\gamma_X < 0$)) state, for increasing and
finite $\Delta m$ the value lies somewhat in between and the different
situations can be hardly distinguished. However, as seen in
Fig.~\ref{fig:tsallis-factor} the numerical value $N_{\Delta m} /N_X
\sim 0.5-0.6$ is rather stable for a reasonable range of $B_X$ and
$\sigma$ values. If we re-interpret $N_X$ as $N_{X,\Delta m}$ this falls
remarkably in the bulk of Fig.~\ref{fig:branching} where $N_{X,\Delta
  m}/N_d \sim 0.5$ implies $N_{X}/N_d \sim 1$. Thus, unlike
expectations, we {\it do not} find the production rate to change
dramatically due to binding energy effects due to $\Delta m$; the
$p_T$ shape will likewise not depend on this (unlike
expectations~\cite{Esposito:2015fsa}).  In a recent and insightful
paper Kang and Oller have analyzed the character of the $X(3872)$ in
terms of bound and virtual states within simple analytical
parameterizations~\cite{Kang:2016jxw}. While the Dashen-Kane
cancellation has not been explicitly identified, it would be
interesting to see if their trends can be reproduced by more
microscopic approaches.

\section{Conclusions}

Theoretically, it is appealing an scenario where the $X(3872)$ is a
half-bound state (zero binding energy) corresponding to the so-called
unitarity limit, characterized by scale
invariance~\cite{Baker:1999np}. In this case, the phase-shift becomes
$\delta= \pi/2$ around threshold, and the occupation number becomes a
half of that of the bound state. Our analysis shows that the large
production rate of the $X(3872)$ at finite $p_T$ does not depend
strongly on the details of the binding since the experimental bin size
is much larger than the binding energy. We also find striking and
universal shape similarities with the $\psi(2S)$ and deuteron
production data via a common Tsallis distribution. A more direct check
of our predicted mild suppression might be undertaken if all
production data would be within the same $p_T$ values. Finally, we
note that in order to envisage a clear fingerprint of the $X(3872)$
binding character a substantial improvement on the current
experimental resolution of its decay products would be required.

\end{multicols}

\begin{multicols}{2}
\acknowledgments{One of us (E.R.A.) thanks Airton Deppman for discussions on Tsallis
distributions. This work is partly supported by the Spanish Ministerio
de Econom¨ªa y Competitividad and European FEDER funds (grants
FPA2016-77177-C2-2-P and FIS2017-85053-C2-1-P) and Junta de
Andaluc\'ia (grant FQM-225).}

\end{multicols}

\vspace{10mm}

\vspace{-1mm}
\centerline{\rule{80mm}{0.1pt}}
\vspace{2mm}

\begin{multicols}{2}

\end{multicols}

\clearpage

\end{document}